\documentclass[cits]{PoS}

\usepackage{amsmath}

\newcommand{\ka}{\kappa}
\newcommand{\al}{\alpha}

\newcommand{\de}{\delta} 

\newcommand{\ba}{\begin{align}} 
\newcommand{\ea}{\end{align}}	
\newcommand{\eref}[1]{eq.~(\ref{#1})}
\newcommand{\fref}[1]{fig.~\ref{#1}}
\newcommand{\nnnl}{\nonumber\\}	

\title{Infrared scaling solutions beyond the Landau gauge: The maximally Abelian gauge and Abelian infrared dominance}

\ShortTitle{Maximally Abelian gauge and Abelian infrared dominance}

\author{\speaker{Markus Q. Huber}\\
        Theoretisch-Physikalisches Institut, Friedrich-Schiller-Universit\"at Jena, Max-Wien-Platz 1, 07743 Jena, Germany\\
        E-mail: \email{markus.huber@uni-jena.de}}

\author{Reinhard Alkofer\\
       Institut f\"ur Physik, Karl-Franzens-Universit\"at Graz, Universit\"atsplatz 5, 8010 Graz, Austria\\
       E-mail: \email{reinhard.alkofer@uni-graz.at}}

\author{Kai Schwenzer\\
       Department of Physics, Washington University, St. Louis, MO 63130, USA\\
       E-mail: \email{schwenzer@physics.wustl.edu}}

\abstract{Functional equations like exact renormalization group and Dyson-Schwinger equations have contributed to a better understanding of non-perturbative phenomena in quantum field theories in terms of the underlying Green functions. In Yang-Mills theory especially the Landau gauge has been used, as it is the most accessible gauge for these methods. The growing understanding obtained in this gauge allows to proceed to other gauges in order to obtain more information about the relation of different realizations of the confinement mechanism. In the maximally Abelian gauge first results are very encouraging as a variant of Abelian infrared dominance is found: The Abelian part of the gauge field propagator is enhanced at low momenta and thereby dominates the dynamics in the infrared. Its role is therefore similar to that of the ghost propagator in the Landau gauge, where one denotes the corresponding phenomenon as ghost dominance. Also the ambiguity of two different types of solutions (decoupling and scaling) exists in both gauges. Here we present how the two solutions are related in the maximally Abelian gauge. The intricacy of the system of functional equations in this gauge required the development of some new tools and methods as, for example, the automated derivation of the equations by the program \textit{DoFun}. We also present results for linear covariant and ghost anti-ghost symmetric gauges.}

\FullConference{The many faces of QCD\\
		November 2-5, 2010\\
		Gent Belgium}

\begin{document}

\section{Dual superconductor and Abelian infrared dominance}

Despite the success of quantum chromodynamics in describing many aspects of the strong interaction we still lack a complete understanding of the phenomenon of confinement in terms of the underlying fundamental fields, the quarks and gluons. Many scenarios have been brought forward, among these the dual superconductor picture \cite{Mandelstam:1974pi,'tHooft:1975pu}. It explains the absence of quarks from the physical spectrum by an analog to the Mei\ss ner-Ochsenfeld effect of solid state physics: The vacuum contains condensed chromomagnetic monopoles\footnote{Although magnetic monopoles are gauge independent objects \cite{Bonati:2010tz} their detection in lattice calculations depends on the choice of gauge \cite{Bonati:2010bb}.} which squeeze the chromoelectric field lines between two quarks into a flux tube. These vortex like structures can be identified best after reducing the gauge symmetry from the non-Abelian $SU(N)$ to the Abelian $U(1)^{N-1}$. Several such partial gauge fixings are known \cite{'tHooft:1981ht}. The most prominent one is the maximally Abelian gauge (MAG), which is defined by minimizing the components of the gauge field which are not Abelian. Since the corresponding generators of the algebra are off-diagonal matrices, these field components are also called \textit{off-diagonal gluon fields} in contrast to the \textit{diagonal} ones, which correspond to the Abelian part of the gluons.

The dual superconductor picture suggests that for the infrared (IR) properties of Yang-Mills theory the diagonal gluons play a dominant role, since the monopoles live in the Cartan subalgebra \cite{Ezawa:1982bf}. This idea was formulated by Ezawa and Iwazaki as the hypothesis of Abelian IR dominance. In their original paper \cite{Ezawa:1982bf} the exact mechanism of this Abelian dominance was not specified any further. In terms of the Green functions one possible realization is a massive behavior for all three propagators (diagonal and off-diagonal gluons and ghosts), but with the lowest mass for the diagonal gluon such that the other fields effectively decouple and the IR dynamics is dominated by the diagonal field. Such a behavior was found by lattice calculations \cite{Bornyakov:2003ee,Mendes:2008ux} and in the refined Gribov-Zwanziger framework \cite{Capri:2008ak,Capri:2010an}. Another possible realization of Abelian IR dominance is a diverging diagonal gluon propagator. Such a behavior was found in an IR scaling analysis of the MAG \cite{Huber:2009wh,Huber:2010ne}. We want to stress that the coexistence of both solutions is not unlikely. Indeed, such a scenario is also known from the Coulomb \cite{Szczepaniak:2001rg,Epple:2007ut} and the Landau gauges \cite{Boucaud:2008ji,Fischer:2008uz}.

We will recall the definition of the MAG in Section \ref{sec:MAG} and recapitulate the cornerstones of an IR scaling analysis in Section \ref{sec:scalingAnalysis}, where also the results for the MAG are presented. Additionally we provide new numerical data corroborating the possible existence of the scaling solution in the MAG and explain how the found scaling solution might be related to the results from lattice calculations. In Section \ref{sec:otherGauges} we make a short detour to linear covariant and ghost anti-ghost symmetric gauges.

\section{The maximally Abelian gauge}
\label{sec:MAG}

For the MAG we split the gauge field into diagonal and off-diagonal components as follows:
\begin{align}
 A_\mu=T^i A^i_\mu+T^a B^a_\mu,
\end{align}
where the $T^i$ are the $N-1$ generators of the Cartan subalgebra of $SU(N)$, i.e., $[T^i,T^j]=0$. For example, in $SU(3)$ the diagonal generators are $T^3$ and $T^8$. A widespread convention is to use the indices $i$, $j$ and so on for diagonal values only and $a$, $b$ and so on for off-diagonal ones. Furthermore we use $A$ for the diagonal and $B$ for the off-diagonal gluon fields. This splitting has interesting consequences for the possible interactions, which can be inferred from the standard commutation relation of the generators:
\begin{align}\label{eq:structure-constants}
 [T^r,T^s]=i\,f^{rst} T^t,
\end{align}
where $r$, $s$ and $t$ can be either diagonal or off-diagonal. From \eref{eq:structure-constants} we can directly see that only three 
off-diagonal fields
or two off-diagonal and one diagonal fields can interact. Even more, as in $SU(2)$ there are only two off-diagonal generators the first possibility also drops out. Thus we have a higher number of interactions in $SU(N>2)$ than in $SU(2)$. In summary the pure Yang-Mills part has the following interactions: $ABB$, $AABB$, $BBBB$, ($BBB$, $ABBB$). The two interactions in parentheses vanish for $SU(2)$.

The idea of the MAG is to minimize the norm of the off-diagonal gluon fields so as to stress the role of the diagonal part. The condition
\begin{align}
 R_{MAG}=\frac1{2} \int dx \, B^a_\mu(x) B^a_\mu(x) \quad \rightarrow \quad \min
\end{align}
is fulfilled if $D^{ab}_\mu B_\mu^b=0$, where the 
respective covariant derivative
contains only the diagonal gluon field:
\begin{align}
D_\mu^{ab}:=&\delta^{ab}\partial_\mu+g\,f^{abi} A_\mu^i.
\end{align}
Since this gauge fixing condition is non-linear the resulting Faddeev-Popov terms in the action contain quartic terms.
The remaining $U(1)^{N-1}$ symmetry of the action is fixed to the Landau gauge, $\partial_\mu A_\mu^i=0$.
Since the field $A^i$ is Abelian the corresponding ghosts decouple so that only the ghosts of the non-diagonal sector remain. Finally, a quartic ghost term is added to the action for renormalizability \cite{Min:1985bx}. The interactions stemming from the gauge fixing and the renormalizability requirement are $Acc$, $AAcc$, $BBcc$, $cccc$, ($Bcc$, $ABcc$). The two interactions in parentheses vanish for $SU(2)$.

The complete resulting action is
\begin{align}
S_{MAG}&= \int dx \Big(\frac{1}{4}F_{\mu\nu}^i F_{\mu\nu}^i+\frac{1}{4}F_{\mu\nu}^a F_{\mu\nu}^a +\bar{c}^a \hat{D}_\mu^{ab}D_\mu^{bc} c^c-g\,f^{bcd}\bar{c}^a \hat{D}_\mu^{ab} B_\mu^c c^d -g^2\,\zeta f^{abi}f^{cdi} B_\mu^b B_\mu^c \bar{c}^a c^d-\nnnl
&-\frac1{2\alpha}(\hat{D}_\mu^{ab} B_\mu^b)^2+\frac{\alpha}{8} g^2 f^{abc}f^{ade} \bar{c}^b c^c \bar{c}^d c^e-\frac{1}{2} g\,f^{abc} (\hat{D}_\mu^{ad} B_\mu^d) \bar{c}^b c^c +\nnnl
&+\frac1{4}g^2\alpha f^{abi}f^{cdi}\bar{c}^a\bar{c}^b c^c c^d+\alpha\frac1{8}g^2 f^{abc}f^{ade} \bar{c}^b \bar{c}^c c^d c^e+\frac1{2\xi}(\partial_\mu A_\mu^i)^2\Big),
\end{align} 
where $\alpha$ is the gauge fixing parameter of the non-diagonal part and $\xi$ that of the diagonal one. Since $\al$ appears in the denominator of vertices it cannot, in contrast to the Landau gauge, be set to zero right away. The dependence on this parameter will be further elaborated in Section \ref{sec:scalingAnalysis}.

\begin{figure}[tb]
\begin{center}
 \includegraphics[width=0.92\textwidth]{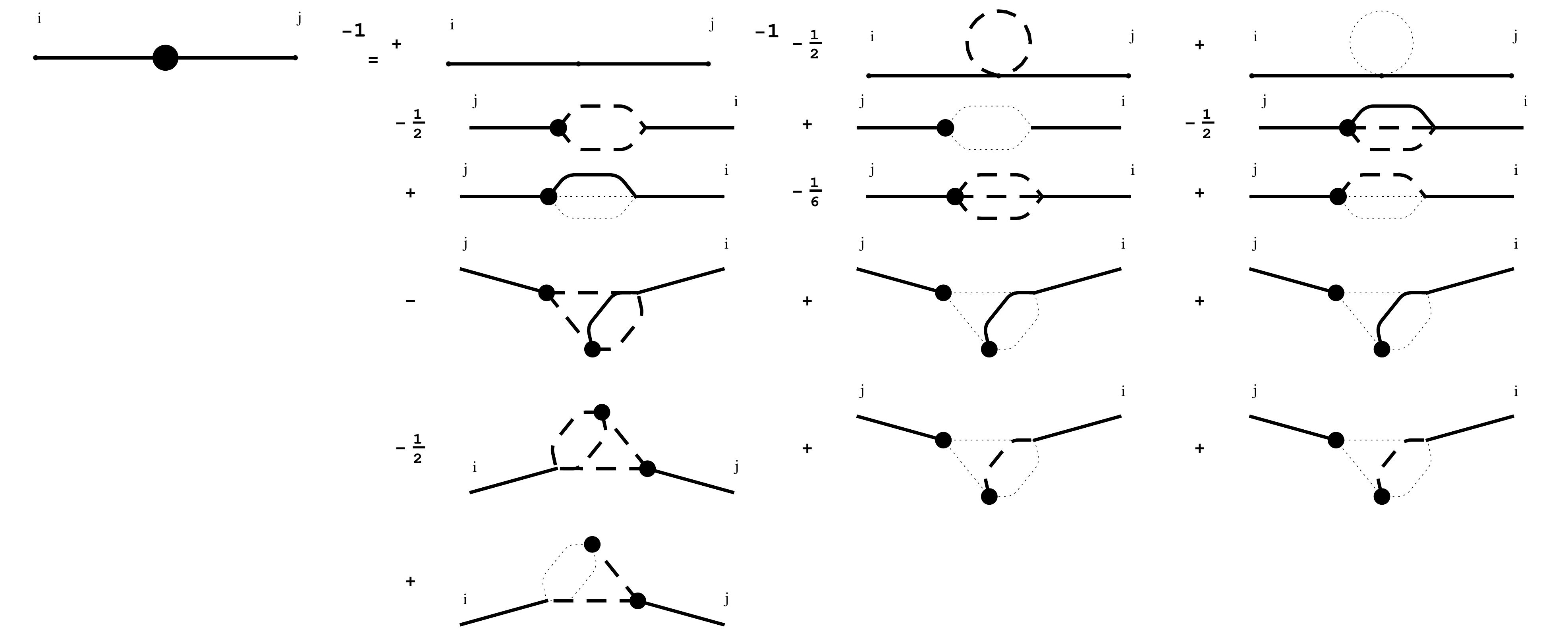}
\caption{\label{fig:MAG-DSEs-AA} DSE of the diagonal gluon two-point function obtained with \textit{DoFun} \cite{Huber:2011qr,Alkofer:2008nt}. Continuous lines denote diagonal gluons, dashed ones off-diagonal gluons and dotted ones ghosts.}
\end{center} 
\end{figure}

Since the MAG has three fields with eleven interactions the derivation of its Dyson-Schwinger equations (DSEs) or functional renormalization group equations (FRGEs) is very cumbersome when doing manually.
To this end the program \textit{DoFun} \cite{Huber:2011qr} has been developed which
can exactly do this for us and the derivation process is considerably shortened. In \fref{fig:MAG-DSEs-AA} the DSE of the gluon two-point function is shown as an example. \textit{DoFun} is the successor of \textit{DoDSE} \cite{Alkofer:2008nt}. Two new features are the inclusion of FRGEs and the derivation of the Feynman rules from a given action. Furthermore the complete algebraic form of the integrands can now be obtained so that computations directly with \textit{Mathematica} are possible. This was very helpful for the calculations of the IR leading diagrams of the MAG \cite{Huber:2010ne} and the Gribov-Zwanziger action \cite{Huber:2009tx}. For the latter the application of \textit{DoFun} helped to identify a unique solution \cite{Huber:2010cq}: Only with the help of a computer algebra system 
we could show that one of the two possible scaling solutions does not yield a numerical solution for the infrared exponent and can thus not be realized.

\section{Scaling analysis}
\label{sec:scalingAnalysis}

Since functional equations are formulated in terms of non-perturbative Green functions they are well suited for the investigation of the IR regime of Yang-Mills theories \cite{Alkofer:2000wg,Fischer:2006ub}. The gauge which received the most attention is without doubt the Landau gauge. Here two solutions have emerged from the analysis of the system of gluon and ghost propagator DSEs: The scaling solution \cite{Fischer:2008uz,Huber:2009tx,Alkofer:2000wg,Fischer:2006ub,vonSmekal:1997vx,Pawlowski:2003hq,Lerche:2002ep,Alkofer:2008jy,Alkofer:2004it} and the decoupling solution \cite{Boucaud:2008ji,Fischer:2008uz,Alkofer:2008jy,Dudal:2007cw,Aguilar:2008xm}. The former is characterized by an IR vanishing gluon propagator and an IR enhanced ghost propagator, whereas the latter has an IR finite gluon propagator and a finite ghost dressing function. Both solutions are connected by the choice of a boundary condition required to solve the system of equations \cite{Boucaud:2008ji,Fischer:2008uz}. Whereas such a choice might seem arbitrary at first sight, it turns out that it only affects the deep IR behavior of the propagators, while all solutions agree for higher momenta. Up to now no calculation exists that shows a difference between these two solutions for physical quantities: In ref. \cite{Braun:2007bx} it was shown that both solutions yield a confining Polyakov loop potential, the confinement-deconfinement and the chiral transition temperatures were found to be the same for both solutions \cite{Fischer:2009gk} and even meson masses are unaffected  by choosing one of the two \cite{Blank:2010pa}. Consequently, the interpretation of this boundary condition as an additional gauge fixing parameter as in the Landau-$B$ gauges of ref. \cite{Maas:2009se} seems reasonable.
We will see that for the MAG we also have two qualitatively different solutions and a similar connection between them exists.

For the scaling analysis we make the ansatz that all dressing functions are characterized by a power law in the IR, e.g., for a propagator we have $D(p)=c(p^2)/p^2$ and $c^{IR}=d\cdot(p^2)^{\de}$, where $\de$ is the so-called infrared exponent (IRE). As we are only interested in the qualitative behavior of the Green functions we focus on the different IREs. Indeed, it is even possible to shift the analysis to the level of the IREs, because all integrals are dominated by low momenta if the external momenta are low and we can replace all propagators and vertices by the corresponding IR expressions \cite{Alkofer:2004it,Huber:2007kc}. The scaling of such an expression can then be determined by counting all exponents of momenta. Thus we can determine the IRE of any given diagram.

The best method to derive the scaling relations of the IREs is the combined use of DSEs and FRGEs as first done for the Landau gauge in refs. \cite{Fischer:2006vf,Fischer:2009tn}. During the study of the MAG it turned out that quite generically the infinitely many relations can be reduced to a rather small number. For details we refer to refs. \cite{Huber:2009wh,Huber:2010ne} where these general relations are written down. Knowing them even allows to investigate complex actions like the Gribov-Zwanziger action with functional methods \cite{Huber:2009tx}.

For the MAG we parametrize the propagators as ($\xi=0$)
\begin{align}
 D_A^{ij}(p^2)&=\de^{ij}\frac{c_A(p^2)}{p^2}\left(g_{\mu\nu}-\frac{p_\mu p_\nu}{p^2}\right),\\
 D_B^{ab}(p^2)&=\de^{ab}\frac{c_B(p^2)}{p^2}\left(g_{\mu\nu}-(1-\al)\frac{p_\mu p_\nu}{p^2}\right),\\
 D_c^{ab}(p^2)&=-\de^{ab}\frac{c_c(p^2)}{p^2}
\end{align}
and take the following power laws for the dressing functions in the IR:
\begin{align}
 c_A(p^2)&\overset{p^2\rightarrow 0}{=} d_A \cdot (p^2)^{\de_A},&\quad c_B(p^2)&\overset{p^2\rightarrow 0}{=} d_B \cdot (p^2)^{\de_B},&\quad c_c(p^2)&\overset{p^2\rightarrow 0}{=} d_c \cdot (p^2)^{\de_c}.
\end{align}
As we do not set the gauge fixing parameter of the off-diagonal part to zero the longitudinal part of the off-diagonal propagator could have its own dressing function. 
In order to take that into account one could introduce
an extra IRE $\de_{B,long}$ and split the off-diagonal field in transverse and longitudinal parts.
However, inserting the bare vertices explicitly and projecting the DSE of the off-diagonal
propagator transversely and longitudinally one finds that the equations for both parts are
the same. Thus the IR analysis yields $\de_{B}=\de_{B,long}$. This result can only be invalidated
if cancellations between different diagrams occurred or due to contributions from the unknown
dressed vertices. Based on this relation we adopt in the following only one common
dressing function for both tensors.

The only consistent scaling relation which can be obtained for the MAG is \cite{Huber:2009wh}
\begin{align}\label{eq:MAGScalingRelation}
 \ka_{MAG}:=-\de_A=\de_B=\de_c\geq0.
\end{align}
This means that the diagonal degrees of freedom are IR enhanced and the off-diagonal ones are IR suppressed. An upper bound for the parameter $\ka_{MAG}$ can be obtained from demanding that the Fourier transformations of the propagators should exist \cite{Lerche:2002ep}: $\ka_{MAG}<1$. The IREs for the vertices can also be obtained \cite{Huber:2009wh}. It turns out that vertices become more IR divergent when the number of off-diagonal legs increases.
The method developed in \cite{Huber:2009wh} has been applied to the MAG in $SU(2)$ and $SU(N>2)$. Interestingly, no differences have been found. The additional interaction of the latter do neither invalidate the $SU(2)$ solution nor do they allow an additional one.

In the derivation of this scaling solution the bare two-point functions play a special role in the respective DSEs. In order to allow for a scaling solution at least one of them has to drop out at zero momentum. This is well known for the Landau gauge, where the bare ghost two-point function is cancelled when choosing the corresponding renormalization condition \cite{Fischer:2008uz,vonSmekal:1997vx,Zwanziger:2001kw}. As we do not know at the beginning which function could be IR divergent, we allow all of them to become IR divergent. At the end it turns out that the only consistent solution is the one with an IR enhanced diagonal gluon propagator. This requirement immediately opens the possibility to check whether the value of the diagonal two-point function at zero momentum can serve as an additional gauge fixing parameter similar to the ghost two-point function in the Landau-$B$ gauges \cite{Maas:2009se}. Indeed, when we take an IR finite diagonal gluon propagator we find that the off-diagonal propagators become IR finite too. The reason is the presence of quartic interactions between diagonal and off-diagonal fields of the MAG. If the diagonal gluon field is massive, the tadpole diagrams in the DSEs originating from these interactions are proportional to the diagonal gluon mass. Consequently the left-hand side should be constant for vanishing momentum from which we conclude that the off-diagonal propagators also become IR finite. This is the mechanism how the zero momentum value of the diagonal gluon propagator switches between the scaling and the decoupling types of solutions.
Note that there is no quartic interaction between the ghost and the gluon in the Landau gauge and consequently an IR finite gluon propagator does not entail an IR finite ghost propagator. 
   
The result of the scaling analysis described above is of course not sufficient evidence that this solution really exists. A full-fledged numerical calculation, on the other hand, is complicated by the structure of the scaling solution of the MAG: The IR leading diagrams are all two-loop. Consequently, for a consistent numerical treatment a truncation scheme rather different from the one used in the Landau gauge has to be found. In the latter it is possible to neglect the two-loop diagrams and still have a truncation consistent in the IR and the UV, because the leading diagrams in both regimes are one-loop. 

\begin{figure}[tb]
\begin{center}
\includegraphics[width=0.45\textwidth]{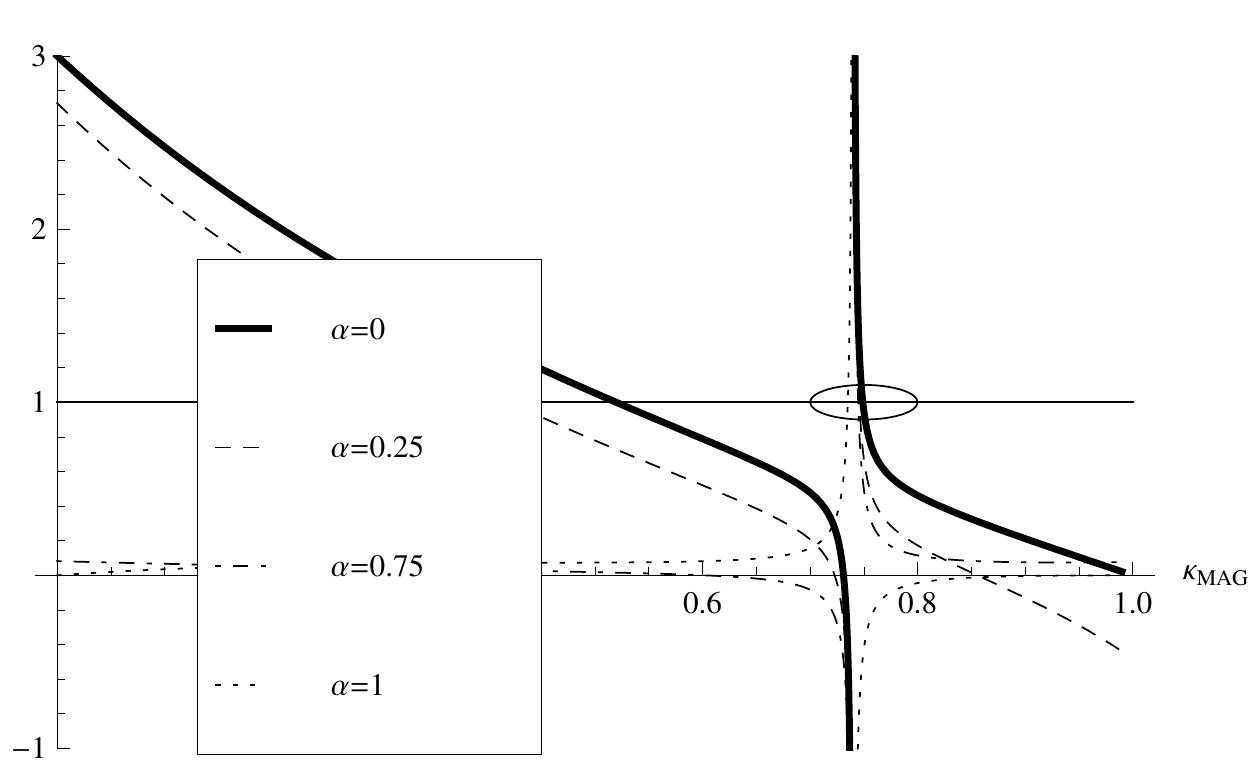}
\includegraphics[width=0.45\textwidth]{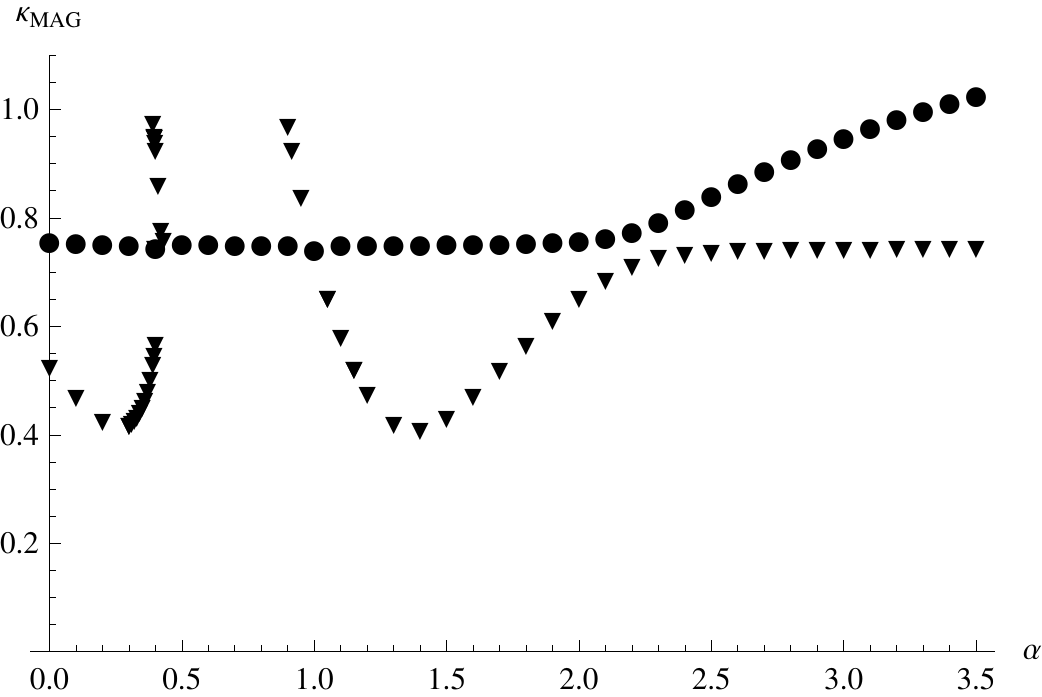} 
\caption{\label{fig:kappaMAG}On the left-hand side the evaluation of eq. (3.9) is shown for several values of the gauge fixing parameter $\alpha$. Crossings with the horizontal line represent solutions for $\ka_{MAG}$. On the right-hand side the two solution branches for varying $\alpha$ are shown.}
\end{center} 
\end{figure}

As a preliminary step towards a numerical (scaling) solution in the MAG we calculated a value for the parameter $\ka_{MAG}$. For this one only needs the IR leading diagrams. Here we find again an ambiguity, since it is not clear whether the squint diagrams are IR leading \cite{Huber:2009wh,Huber:2010ne}. We only took the sunset diagrams, which are definitely IR leading, so that the properly projected equations reduce in the IR to
\begin{align}
 d_A^{-1}&=-X^A_{AABB}(p^2,\ka_{MAG}) d_A d_B^2 - X^A_{AAcc}(p^2,\ka_{MAG})  d_A d_c^2,\\
 d_B^{-1}&=-X^B_{AABB}(p^2,\ka_{MAG})  d_A^2 d_B,\\
 d_c^{-1}&=-X^c_{AAcc}(p^2,\ka_{MAG})  d_A^2 d_c.
\end{align}
The $X(p^2,\ka_{MAG})$ denote the sunset integrals without the coefficients from the propagator power laws. The superscript indicates the corresponding DSE and the subscript the bare vertex contained in the diagram. Using the invariant combinations $ I_1:=d_A^2 d_B^2$ and $I_2:=d_A^2 d_c^2$ the three equations can be combined to
\begin{align}\label{eq:eqForKappa}
 1= \frac{X^A_{AABB}(p^2,\ka_{MAG})}{ X^B_{AABB}(p^2,\ka_{MAG})} + \frac{X^A_{AAcc}(p^2,\ka_{MAG})}{X^c_{AAcc}(p^2,\ka_{MAG})}.
\end{align}
If the $X(p^2,\ka_{MAG})$ are known, this equation yields the solution(s) for $\ka_{MAG}$. The computation of the integrals requires the dressed four-point functions. Their power law behavior is known (they are IR constant), but their tensor structures are not. As simple ans\"atze we employed the tree-level structures\footnote{All tree-level expressions can be found in \cite{Huber:2010ne}.}. The last unspecified quantity is the gauge fixing parameter of the off-diagonal part, $\alpha$. In the left plot of \fref{fig:kappaMAG} we show \eref{eq:eqForKappa} for several values of $\al$. Surprisingly there are solutions $0.7<\ka_{MAG}<0.8$ for all tested values of $\al$. The plot on the right-hand side underlines this: For certain ranges of $\alpha$ there are two solution branches, but there is always one around $\ka_{MAG}\approx 0.74$. The stability with respect to changes of $\alpha$ is by no means required and even persists for rather large values of $\alpha$. This is in big contrast to linear covariant gauges, where the existence of a naive scaling solution in accordance with the Slavnov-Taylor identities is only assured for $\xi=0$, see Section \ref{sec:otherGauges}.
We want to stress the fact that finding a sensible solution for $\ka_{MAG}$ is a non-trivial result: The scaling analysis itself does not rely on any details of the vertices except for their combinatorial properties, while the calculation of $\ka_{MAG}$ depends on all the little details such as Lorentz and color structure. In this calculation the program \textit{DoFun} \cite{Alkofer:2008nt,Huber:2011qr} was of great help, since the intermediate expressions of the calculation would cover several pages.
 
\section{Other gauges}
\label{sec:otherGauges}

The scaling analysis can be also be employed to other gauges. Here we present the results for linear covariant and ghost anti-ghost symmetric gauges, which generalize the analysis of ref. \cite{Alkofer:2003jr}, where bare vertices were employed, to the case of dressed vertices.

The decisive difference between the MAG and the linear covariant gauges is that the longitudinal part of the gluon propagator does not acquire a dressing in the latter but stays proportional to the gauge fixing parameter $\bar{\xi}$:
\begin{align}\label{eq:gluonPropLinCov}
 D_{A,\mu\nu}^{rs}(p^2)=\de^{rs}\left(g_{\mu\nu}-\frac{p_\mu p_\nu}{p^2} \right) \frac{c_A(p^2)}{p^2}+\bar{\xi} \frac{p_\mu p_\nu}{p^4}.
\end{align}
This can be inferred either from a Slavnov-Taylor identity \cite{Alkofer:2000wg} or from the fact that the longitudinal part of the gluon propagator, $\langle (\partial_\mu A_\mu) (\partial_\nu A_\nu) \rangle$, corresponds to the second moment of the Gaussian distribution, $e^{-\frac{1}{2\xi}(\partial_\mu A_\mu)^2}$, and is hence proportional to $\xi$ \cite{Cucchieri:2008zx}.
To include the longitudinal part of the gluon into the present formalism we can interpret it as an additional field. Its propagator has the IRE $\de_{A,long}$. This allows to distinguish also between the transverse and longitudinal parts of the vertices. Naively the same inequalities are obtained for the longitudinal gluon field as for the transverse one.
This could change if some vertices vanished when they are contracted with a longitudinal gluon propagator. After a more careful analysis one finds that indeed some diagrams drop out, but at the end those turn out not to be important. 
The vertex which is decisive for the IR leading behaviour
is again the ghost-gluon vertex and we obtain for the longitudinal gluon propagator IRE the same result as for the transverse one, as the longitudinal part of the ghost-gluon vertex does not vanish. The result $\de_A=\de_{A,long}=-2\de_c$ leads together with the triviality of the longitudinal part, $\de_{A,long}=0$, to $\de_A=-2\de_c=0$ \cite{Huber:2009wh,Huber:2010ne}. This allows two possible conclusions: Either there is no non-trivial scaling relation for linear covariant gauges, or the naive application of the IR analysis is insufficient here. 

Finally we consider ghost anti-ghost symmetric gauges. Their most important difference to the Landau gauge is that they feature an additional quartic ghost interaction. This interaction is exactly what denies a consistent non-trivial scaling solution for this gauge \cite{Huber:2009wh,Huber:2010ne}: As in linear covariant gauges we obtain $\de_A=\de_c=0$, i.e., a trivial scaling solution.

\section{Summary}

The MAG is a convenient choice of gauge for the investigation of the hypothesis of Abelian IR dominance. In addition to the decoupling type solution already known from lattice simulations and the refined Gribov-Zwanziger framework our results hint at the existence of a scaling type solution where the diagonal gluon propagator is IR dominant and the off-diagonal degrees of freedom are IR suppressed. The IREs for the complete tower of Green functions can be derived in this case. The two types of solutions are related via the zero momentum value of the diagonal gluon two-point function: If it is chosen as zero the scaling solution is obtained, whereas we could show that a finite value at zero momentum automatically leads to finite values at zero momentum for the off-diagonal propagators, i.e., the decoupling type solution is realized. If the situation is similar to the Landau gauge, this choice is a choice of gauge and should not have any consequences for physical observables. A common feature of both solutions is that the diagonal degrees of freedom dominate in the IR in support of the Abelian IR dominance hypothesis. Thus further investigations using the MAG are promising to obtain a better understanding of the confinement mechanism. Finally we want to stress that the MAG is the first gauge besides the Landau gauge where a full scaling solution was found. We investigated also linear covariant and ghost anti-ghost symmetric gauges, but straightforward scaling solutions do not seem possible and more detailed analyses are required.

\acknowledgments
We thank the organizers of the workshop for providing such a pleasant and stimulating environment where so many interesting and fruitful discussions were possible. 
M.Q.H. was supported by the DFG project GRK 1523 and the FWF project W1203.

\bibliographystyle{utphys_mod}
\bibliography{literature_MAG}

\end{document}